# Enhanced Valley Zeeman Splitting in Fe-Doped Monolayer MoS$_2$

Qi Li[1,8], Xiaoxu Zhao[2,8], Longjiang Deng[1], Zhongtai Shi[1], Sheng Liu[4], Qilin Wei[5], Linbo Zhang[1], Yingchun Cheng[5*], Li Zhang[1], Haipeng Lu[1], Weibo Gao[4], Wei Huang[5], Cheng-Wei Qiu[7], Gang Xiang[6], Stephen John Pennycook[2], Qihua Xiong[4], Kian Ping Loh[3*] and Bo Peng[1*]

[1]National Engineering Research Center of Electromagnetic Radiation Control Materials, School of Electronic Science and Engineering, University of Electronic Science and Technology of China, Chengdu, 611731, China

[2]Department of Materials Science and Engineering, National University of Singapore, 9 Engineering Drive 1, 117575, Singapore

[3]Department of Chemistry and Centre for Advanced 2D Materials and Graphene Research Centre, National University of Singapore, 117549, Singapore

[4]Division of Physics and Applied Physics, School of Physical and Mathematical Sciences, Nanyang Technological University, 637371, Singapore

[5]Key Laboratory of Flexible Electronics & Institute of Advanced Materials, Nanjing Tech University, Nanjing, 211816, China

[6]College of Physical Science and Technology, Sichuan University, Chengdu, 610064, China

[7]Department of Electrical and Computer Engineering, National University of Singapore, 117583, Singapore

[8]Qi Li and Xiaoxu Zhao contributed equally to this work.

*Correspondence to bo_peng@uestc.edu.cn (B.P.), iamyccheng@njtech.edu.cn (YC.C), chmlohkp@nus.edu.sg (K.P.L)

**Abstract:** The "Zeeman effect" offers unique opportunities for magnetic manipulation of the spin degree of freedom (DOF). Recently, valley Zeeman splitting, referring to the lifting of valley degeneracy, has been demonstrated in two-dimensional transition metal dichalcogenides (TMDs) at liquid helium temperature. However, to realize the practical applications of valley pseudospins, the valley DOF must be controllable by a magnetic field at room temperature, which remains a significant challenge. Magnetic doping in TMDs can enhance the Zeeman splitting, however, to achieve this experimentally is not easy. Here, we report unambiguous magnetic manipulation of valley Zeeman splitting at 300 K ($g_{eff}$= -6.4) and 10 K ($g_{eff}$= -11) in a CVD-grown Fe-doped $MoS_2$ monolayer; the effective Landé $g_{eff}$ factor can be tuned to -20.7 by increasing the Fe dopant concentration, which represents an approximately fivefold enhancement as compared to undoped $MoS_2$. Our measurements and calculations reveal that the enhanced splitting and $g_{eff}$ factors are due to the Heisenberg exchange interaction of the localized magnetic moments (Fe 3$d$ electrons) with $MoS_2$ through the $d$-orbital hybridization.

**Keywords:** magnetic doping, valleytronic, spin, ferromagnetic, 2D transition metal dichalcogenides

Room-temperature manipulation of the spin degree of freedom (DOF) is crucial for spintronic and valleytronic devices.[1, 2] Two-dimensional transition metal dichalcogenides (TMDs) exhibit a distinguishable carrier distribution in two momentum-dependent valleys;[3-7] the broken spatial inversion symmetry together with the strong spin-orbit coupling (SOC) results in the coupling of the spin and valley, giving rise to valley-dependent optical selection rules.[8] The spin-up (-down) holes and spin-down (-up) electrons mostly exist in the $+K$ ($-K$) momentum valley under optical pumping by the right-handed $\sigma^+$ (left-handed $\sigma^-$) polarized light. The spin and valley ($+K$ and $-K$) information can be optically detected with polarization-resolved photoluminescence (PL). The chiral optical selection rules allow that valley Zeeman splitting ($\Delta E$) can be read-out optically,[9] which is defined as the energy difference between the opposite circularly polarized PL, $\Delta E = E(\sigma^-) - E(\sigma^+) = 2(\Delta_L + \Delta_V + \Delta_S)$. The total cumulative magnetic moment $\mu$ consists of atomic orbital magnetic moment $\mu_L$ ($m\mu_B$), valley magnetic moment $\mu_V$ ($\Delta\alpha\tau\mu_B$) and spin magnetic moment $\mu_S$ ($\mu_B$), where $\tau = \pm 1$ represents the index for the $\pm K$ valleys, $m$ is the magnetic quantum number of $d$-orbitals, $\Delta\alpha$ is the difference in the valley $g$-factors of the conduction and valence bands, and $\mu_B$ is the Bohr magneton.[5] The interaction of the magnetic moment $\mu$ with a magnetic field $B$ can lift the valley degeneracy and slightly shift the energy of the valence and conduction bands in the two valleys, resulting in the splitting of the PL spectrum lines of the $+K$ and $-K$ valleys, analogous to the "Zeeman effect". Zeeman-type valley splitting has been studied in monolayer WSe$_2$,[4, 5, 10] monolayer MoS$_2$ and WS$_2$,[11, 12] monolayer MoSe$_2$,[3, 13] and bilayer MoTe$_2$[14] under an out-of-plane magnetic field. The interfacial magnetic exchange field has also been utilized to enhance the Zeeman-type valley splitting in a heterojunction geometry comprising monolayer TMDs and magnetic films (EuS, CrI$_3$).[15, 16]

Increasing the valley Zeeman splitting is of great importance for valley polarization control for logic applications.[17] A 15 meV valley Zeeman splitting was observed using an extremely high magnetic field up to 65 T at 4 K.[11] Considerable efforts have also been devoted to enlarging the effective Landé $g_{eff}$ factor, which is estimated to be approximately -4 in pristine monolayer TMDs.[3, 10, 15, 18] Recently, the $g_{eff}$ factor was observed to be as high as -18 by increasing the electron density in heavily gated monolayer TMDs, which originates from the exciton-polaron resonance and phase-space filling effect.[19-22] In the field of magnetically doped semiconductors, J. K. Furdyna and H. Ohno have done pioneering researches and brought diluted magnetic semiconductor (DMS) to prominence on the world scale.[23-25] The presence of magnetic ions results in the *sp-d* exchange interaction between the band electrons and the localized magnetic moments,[26-28] which further leads to extremely large Zeeman splitting and ferromagnetism.[29-31] Theoretically, magnetic doping in monolayer TMDs has been predicted to increase the magnetic susceptibility, even possibly giving rise to magnetic order and rich physics due to the interaction between the SOC and local ferromagnetism.[32, 33] However, substitutional doping of monolayer TMDs without collateral damage to the lattice is currently very challenging.

In this work, we observed enhanced valley Zeeman splitting in CVD-grown Fe-doped monolayer MoS$_2$ by circularly polarized magneto-photoluminescence, with a $g_{eff}$ factor of -6.4 at 300 K. The $g_{eff}$ factor exponentially increases to approximately -11 as the temperature decreases to 10 K and is further monotonically tuned to -20.7 by increasing the Fe dopant concentration. This value corresponds to an approximately fivefold enhancement as compared to undoped monolayer $MoS_2$. Using first-principles calculations, we propose that the enhanced valley Zeeman splitting and $g_{eff}$ factor are attributed to the isotropic Heisenberg exchange interaction of magnetic moments arising from the *d*-orbital hybridization of Fe and Mo atoms.

**Results and discussion**

**Structure and phase of Fe-doped monolayer $MoS_2$.** Fe-doped monolayer $MoS_2$ was grown on $Si/SiO_2$ substrates by the CVD method (see Methods and Fig. 1a-d). The layer identification was corroborated by direct topography measurements with an atomic force microscope (AFM). The 0.9-nm step height exactly corresponds to the thickness of CVD-grown monolayer $MoS_2$[34, 35] (see Supplementary Information Fig. S1). The in-plane $E_{2g}^1$ vibration mode and out-of-plane $A_{1g}$ vibration mode of CVD-grown monolayer $MoS_2$ (P1) are at ~380.8 and ~401.6 cm$^{-1}$, respectively (Fig. 1e); the energy difference between the $A_{1g}$ and $E_{2g}^1$ modes is ~20.8 cm$^{-1}$, which is consistent with the fingerprint of monolayer $MoS_2$ grown by the CVD method.[36, 37] The Fe atom substitutions induce stiffening of the $E_{2g}^1$ mode and softening of the $A_{1g}$ mode, resulting in $E_{2g}^1$ ($A_{1g}$) shifting to higher (lower) frequency; however, the defects in $MoS_2$ can lead to broadening and an opposite shifts of the $E_{2g}^1$ and $A_{1g}$ Raman peaks.[38] Thus, the shifts of the $A_{1g}$ and $E_{2g}^1$ peaks depend on competing effects by substitutions and disorder in the CVD-grown Fe-doped monolayer $MoS_2$. Two pronounced exciton emission features at ~1.82 and ~1.95 eV are clearly observed at room temperature, which are associated with the A and B excitons of monolayer $MoS_2$;[8, 39] the Fe dopants introduce an additional nonradiative recombination way, resulting in an increasing nonradiative decay rate, thus the PL intensity decreases as the Fe dopant concentrations increase (Fig. 1f).[40]

To validate the substitution of Mo by Fe atoms ($Fe_{Mo}$), we investigated the structure and phase of the CVD-grown Fe-doped $MoS_2$ monolayer on an atomic scale by employing an aberration-corrected scanning transmission electron microscope with an annular dark field detector (STEM-ADF) operated at a low accelerating voltage (60 kV). The maximum energy from 60 kV electron beam is smaller than the threshold to damage of $MoS_2$. The STEM-ADF image represents the inherent elastically scattered electrons; therefore, the image contrast of the ADF image depends on the atomic number $Z$, varying as approximately $Z^{1.6-1.7}$. This phenomenon renders STEM-ADF imaging a powerful tool to unveil the atomic structures of 2D materials.[41] The presence of substitutional doping of Mo by Fe is reflected by the contrast reduction in the STEM-ADF image (Fig. 2a and Supplementary Information Fig. S2). The location and density of Fe dopants were calculated based on the contrast reduction, and the mapping of all atom species of the Fe-doped $MoS_2$ monolayer is depicted in Fig. 2b. Statistical sampling shows that the concentration of Fe substitutes is ~2% based on statistical analysis, suggesting a $Mo_{0.98}Fe_{0.02}S_2$ chemical stoichiometry. The contrast of the Fe atom is approximately equal to that of the S dimer (Fig. 2c), which is confirmed by the corresponding simulated image (Fig. 2d-e). Furthermore, the presence of the Fe dopant is verified by the electron energy loss spectrum (EELs) (Fig. 2f and Supplementary Information Fig. S3); the sharp Fe L-edge features are clearly detected. The presence of Fe dopants (Fig. 2g) does not impact the structural integrity of monolayer $MoS_2$. Neither additional S vacancies nor discernible strain (Fig. 2h) proximal to the Fe dopant site are introduced into the pristine lattices. Therefore, the Fe dopants do not damage the lattice of the $MoS_2$ monolayer, which is attributed to the nearly equal ionic radius between Fe and Mo ions.

**Valley Zeeman splitting in Fe-doped monolayer $MoS_2$.** The valley Zeeman splitting in pristine $MoS_2$ is attributed to the valley magnetic moment and the atomic orbital magnetic moment of the Mo $d$-orbital.[5] However, the introduction of Fe atoms leads to the $d$-orbital hybridization of Fe and Mo atoms, which significantly enhance the valley Zeeman splitting (Fig. 3a). The measurements of polarization-resolved PL spectra under an out-of-plane magnetic field were performed on the Fe-doped monolayer $MoS_2$ using a confocal Raman/PL microscope system, which was checked by polarization-resolved Si Raman spectra obtained under excitation by circularly polarized light at 2.41 and 1.96 eV. The helicity parameters of Si Raman feature are up to 98% and 93%, respectively, verifying that our optical system is excellently calibrated (Supplementary Information Fig. S4). Figure 3b-d shows the normalized polarization-resolved PL spectra of the neutral A exciton of the Fe-doped $MoS_2$ monolayer (P1) at selected values of the magnetic field at 10 K. At 0 T, the $\sigma^-$ circular polarization PL emission from the -$K$ valley (red curve) is completely identical to the $\sigma^+$ PL emission from the +$K$ valley (blue curve). However, at high field, the $\sigma^-$ and $\sigma^+$

PL peaks split due to the time-reversal symmetry breaking; the $\sigma^-$ ($\sigma^+$) PL peak shifts to a higher energy than $\sigma^+$ ($\sigma^-$) at -7 T (+7 T), indicating the occurrence of valley Zeeman splitting. The long-range magnetic orders are forbidden in the Fe-doped MoS$_2$ monolayer at 10 K (see Supplementary Information Fig. S5), thus, the Fe-doped MoS$_2$ monolayer is not ferromagnetic and no valley Zeeman splitting is observed at zero external fields. Valley Zeeman splitting is revealed more clearly through the intensity difference ($\Delta PL$) in the normalized polarization-resolved PL at ±7 T and 0 T (Fig. 3e), $\Delta PL = I(\sigma^-) - I(\sigma^+)$. At zero field (black curve), $\Delta PL$ is almost equal to zero. In contrast, at ±7 T, $\Delta PL$ deviates from zero and displays opposite values symmetric about 0 T when the direction of the magnetic field is switched, with a positive (negative) value for +7 T (-7 T) below ~1.88 eV and an inverse trend above this value. This observation implies that, whereas the magnitude depends on the magnetic field intensity, the sign of valley splitting is determined by the relationship between the helicity and the magnetic field direction. Importantly, valley Zeeman splitting is extracted by computing the "center of mass" of the PL peak. The splitting as a function of $B$ is shown in Fig. 3f, yielding a $g_{eff}$ factor of -11±0.5. The linear magnetic-field-dependence of valley Zeeman splitting in Fe-doped monolayer MoS$_2$ arises from the linear variation of intrinsic Zeeman splitting ($g\mu_B B$) and localized Fe spins ($\langle S \rangle$) with magnetic field.[25, 42]

To further verify the increasing effective $g_{eff}$ factor in Fe-doped monolayer MoS$_2$, exfoliated and undoped CVD-grown MoS$_2$ monolayers were used as control samples and measured at 10 K in the same polarization-resolved optical system (Supplementary Information Fig. S6). The $g_{eff}$ factors of the exfoliated and undoped CVD-grown MoS$_2$ monolayers are -4.4±0.5 and -4.5±0.5 at 10 K, respectively. Thus, the $g_{eff}$ factors of Fe-doped monolayer MoS$_2$ are at least ~250% times larger than that of exfoliated and undoped CVD-grown MoS$_2$ monolayer (Fig. 3f and 3g). The valley Zeeman splitting can be strikingly tuned by controlling the Fe dopant concentration, which significantly increases as the Fe dopant concentration increases. The $g_{eff}$ factor of Fe-doped MoS$_2$ monolayer achieves a maximum of approximately -20.7±1.0, which is ~5 times as larger as that of exfoliated and undoped CVD-grown MoS$_2$ (Fig. 3g and Supplementary Information Fig. S6 and S7).

The absorption spectra of all samples were examined at 10 K (see Supplementary Information Fig. S8a-b). Only excitons A and B are observed, and no distinguishable trion absorption peak due to electron doping is detected.[8] Therefore, the possibility of electron doping leading to the increase in $g_{eff}$ factor is ruled out.[8, 18, 19, 43] We also measured Zeeman-type spin splitting of defect emission in monolayer MoS$_2$ (see Supplementary Information Fig. S8c-f). Strikingly, the $g_{eff}$ factor of defect emission is approximately -6.0 at 10 K, which is only half and even 1/4 of that in Fe-doped MoS$_2$ monolayer (Fig. 3). Thus, the enhancements of $g_{eff}$ factor in Fe-doped MoS$_2$ monolayer are not induced by defects. The Fe dopants are anticipated to lead to the distinct increase in $g_{eff}$ factor and valley Zeeman splitting.

Importantly, unambiguous valley Zeeman splitting is observed at room temperature, although it is smaller than the PL linewidth. At zero field, the σ⁻ and σ⁺ PL spectra from the -$K$ and +$K$ valleys are completely overlapped; however, they split at high field, with inverse shifting directions for +7 T and –7 T (P2, see Supplementary Information Fig. S9). The spectral line shape of $\Delta PL$ at room temperature is consistent with that at 10 K, whereas the $\Delta PL$ intensity decreases by 50%. Thus, $\Delta PL$ clearly shows the occurrence of valley Zeeman splitting at room temperature, which follows a linear relation with $B$, giving a $g_{eff}$ factor of -6.4±0.5 at 300 K (see Supplementary Information Fig. S9a and S9d), which is ~1.6 times as large as that of undoped CVD-grown monolayer $MoS_2$ with a $g_{eff}$ factor of -3.9 at 300 K (see Supplementary Information Fig. S9b and S9d); the $g_{eff}$ factor of Fe-doped $MoS_2$ monolayer also increases with increasing Fe dopant concentration at room temperature, achieving a maximum of approximately -15.7, which represents an enhancement of ~400% as compared to CVD-grown undoped monolayer $MoS_2$ (see Supplementary Information Fig. S9e). Thus, spin manipulation in TMDs can be pushed to room temperature from liquid He temperature limit.

To further understand the valley Zeeman splitting, the temperature dependence of the valley Zeeman splitting and $g_{eff}$ factor were determined on Fe-doped monolayer $MoS_2$. Figure 4a-b shows the polarization-resolved PL spectra as a function of temperature at -7 T and +7 T. The valley Zeeman splitting between the $\sigma^-$ and $\sigma^+$ PL components is distinct, whereas the splitting increases as the temperature decreases from 300 K to 10 K (P1, Fig. 4a-c). Figure 4c shows the valley Zeeman splitting extracted from Fig. 4a-b as a function of temperature at ±7 T, indicating a monotonic increase in the splitting with decreasing temperature. Figure 4d shows the temperature dependence of the effective $g_{eff}$ factor of Fe-doped $MoS_2$ monolayer, which increases by nearly 1 times as the temperature decreases from 300 K to 10 K; in contrast, the $g_{eff}$ factor of undoped CVD-grown MoS$_2$ monolayer almost keep constant and slightly change from -3.9 to -4.5 with the decreasing temperature from 300 K to 10 K (see Supplementary Information Fig. S9d and S6d).

**Heisenberg exchange interaction mechanism for valley Zeeman splitting.** The enhancement of valley Zeeman splitting and $g_{eff}$ factor in Fe-doped monolayer MoS$_2$ can be explained by the *kp* mode using the effective Hamiltonian for the +*K* and –*K* valleys. In undoped monolayer MoS$_2$, the spin splitting in the valleys originates from the inversion symmetry breaking together with the SOC, which are opposite in different valleys due to time-reversal symmetry (Fig. 5a). The total Hamiltonian is expressed as $H(k) = at(\tau k_x \sigma_x + k_y \sigma_y) + \Delta \sigma_z / 2 - \lambda \tau (\sigma_z - 1) s_z / 2$, where *a*, *t*, *Δ*, and *τ* represent the lattice constant, effective hopping integral, energy gap and valley index, respectively, 2λ is the spin splitting at the top of the valence band due to the SOC, and $s_z$ is the Pauli matrix for spin.[44-46] The last term in *H(k)* is the SOC term causing valence band splitting. In the presence of Fe dopant, an additional Heisenberg exchange Hamiltonian term $H_m = -xJ_{eff} \langle S \rangle s_z$ that modifies the valley spin polarization of Fe-doped monolayer MoS$_2$ is introduced, where $x$ and $J_{eff}$ are the dopant concentration, the exchange constant, respectively. $\langle S \rangle$ is the magnetic-field-dependent localized Fe spins along the magnetic field. Thus, the spin splitting at the top of the valence band of the +*K* and –*K* valleys should be expressed as $2\lambda + (2\Delta E_{Fe} + g\mu_B B)/2$ and $2\lambda - (2\Delta E_{Fe} + g\mu_B B)/2$ (Fig. 3a), respectively, where $2\Delta E = xJ_{eff} \langle S \rangle$. Therefore, valley Zeeman splitting is expressed as $\Delta E = g\mu_B B + xJ_{eff}\langle S\rangle$. The Fe dopants result in enhanced valley Zeeman splitting, which increases with increasing dopant concentration. Importantly, previous works have reported that the localized Fe spins $\langle S\rangle$ show a linear variation with magnetic field,[25, 42, 47] which drastically contrasts with the nonlinear dependence of Mn spins in Mn-doped diluted magnetic materials ($Hg_{1-x}Mn_xTe$, $Cd_{1-x}Mn_xSe$, $Cd_{1-x}Mn_xTe$).[28, 29] Therefore, valley Zeeman splitting of monolayer Fe-doped $MoS_2$ is linearly dependent with magnetic field, consistent with the above experimental results, as shown in Fig. 3 and Supplementary Information Fig. S7.

Finally, we briefly discuss the temperature dependence of valley Zeeman splitting and the $g_{eff}$ factor. In the Heisenberg exchange interaction mode, the exchange constant $J_{eff}$ is a sensitive function of the interatomic distance,[48] expressed as $J_{eff} = J_0 \exp(-\chi <r>_T)$, where $<r>_T$ represents the displacement of the equilibrium separation between Mo and Fe atoms from absolute zero ($<r>_0 = 0$), which originates from lattice thermal expansion and displays a positive linear dependence with temperature in $MoS_2$ materials,[49, 50] $<r>_T = \alpha T$. Magnetic order is strongly suppressed by thermal fluctuations in the monolayer Fe-doped $MoS_2$ according to the Mermin-Wagner theorem. Therefore, the temperature dependence of the localized spins originating from Fe atoms is predicted by the Boltzmann distribution ($\exp(-\Delta/k_B T)$) rather than the Brillouin law used in traditional three-dimensional diluted magnetic semiconductor,[42, 51, 52] where $\Delta$ is the energy barriers of Fe spin flipping. Therefore, the valley Zeeman splitting increases with decreasing temperature, leading to an increase in the $g_{eff}$ factor, in agreement with our experimental results (see the solid line in Fig. 4c and 4d). The energy barrier of Fe spin fipping is estimated to be ~$10^{-4}$ meV, indicating that the spin flipping is suppressed only below ~1 mK ($k_BT$), thus, the long-range magnetic order is strongly prohibited and no ferromagnetism is observed in Fe-doped MoS$_2$ monolayer at 10 K (see Supplementary Information Fig. S5), which is consistent with the experimental result of no valley Zeeman splitting at 0 T (Fig. 3c). The lattice thermal expansion dominates the temperature dependence of valley Zeeman splitting and the $g_{eff}$ factor.

Further insights into the origin of the enhanced valley Zeeman splitting can be obtained from the calculated electronic band structures of undoped monolayer MoS$_2$ and Fe-doped monolayer MoS$_2$ (Fig. 5). The calculations show that the formation energy of Fe$_{Mo}$ is lowest in S-rich condition, indicating that the Fe$_{Mo}$ substitutions are favorable (see Supplementary Information Section 10), consistent with our experiments (Fig. 2 and Supplementary Information Fig. S2 and S3). Strikingly, the valley spin states in Fe-doped monolayer MoS$_2$ are different from those in undoped pristine MoS$_2$ monolayer (Fig. 5a-b). Valley Zeeman splitting is clearly observed at and in the vicinity of *K* points in Fe-doped MoS$_2$ monolayer, while there is no any valley splitting in undoped MoS$_2$ monolayer[46]. Figure 5c-e shows the calculated electronic band structures of orbitals of Mo, Fe and S atoms in Fe-doped monolayer MoS$_2$, demonstrating that the states in the valence band of Fe-doped monolayer MoS$_2$ are formed primarily from the *d*-orbitals of Mo and Fe and *p*-orbitals of S. The components of the Fe *d*-orbital are distinct in the electronic band structures, although the *d*-orbital of Mo and *p*-orbitals of S predominates. Figure 5f shows the corresponding orbital percentage of $d_{z^2}$, $d_{x^2-y^2}$ and $d_{xy}$ at the valence band maximum (VBM) and conduction band minimum (CBM) in the $+K$ valley, which dominate the valley Zeeman splitting. The $d$-orbital ratios of Fe to Mo atoms are up to ~1% and ~10% at the CBM and VBM points, respectively, strikingly validating the occurrence of $d$-orbital hybridization of Fe and Mo atoms. The VBM of the Fe-doped MoS$_2$ is mainly composed of the $d_{x^2-y^2}$ and $d_{xy}$ orbitals of Mo and Fe atoms with $m = \pm 2$ in $\pm K$ valley (Fig. 5f-g), while the CBM consists of $d_{z^2}$ orbitals of Mo and Fe atoms with $m = 0$.[53] The density of states (DOS) also verify that the $d$-orbitals of Fe atoms contribute to the orbital hybridization (see Supplementary Information Fig. S11a). Therefore, the Heisenberg exchange interaction of magnetic dopant Fe atoms and MoS$_2$ arises from the hybridization of the $d$-orbitals of Mo and Fe, leading to the enhanced valley Zeeman splitting. In our first-principles calculations, the 4 × 4 (Fig. S10) and 5 × 5 (Fig. 5 and Fig. S11) supercells with one Fe atom replacing one Mo atom were used. The local magnetic moments originating from Fe atoms are ordered at 0 K, analogous to that under a strong external magnetic field. However, the induced effective magnetic field may be much stronger than our applied magnetic field; moreover, the maximum concentration of Fe dopant is only 2% in our experiments, which is much smaller than the doping concentration (~7%) in calculations. Thus, the deviation between the calculated and experimental values of valley Zeeman splitting is reasonable. Nonetheless, the calculated trends are consistent with our experimental observations in supporting the enhanced valley Zeeman splitting in Fe-doped MoS$_2$.

**Conclusion**

In summary, we have demonstrated an enhanced $g_{eff}$ factor and valley Zeeman splitting in Fe-doped monolayer $MoS_2$, arising from the Heisenberg exchange interaction between the local magnetic moments and $MoS_2$ through the *d*-orbital hybridization. This result could potentially give rise to many interesting phenomena such as the tunneling magnetic-resistance effect and spin Hall effect.[54-56] Our proposed approach for magnetic doping and manipulation of valley spin at room temperature is not limited to $MoS_2$ and can be achieved in a variety of TMDs, including $WSe_2$, $MoSe_2$ and $WS_2$. These features highlight the potential to synthesize magnetic 2D materials and to produce quantum confined spintronic and valleytronic devices at room temperature.[57, 58]

**Experimental methods**

**Synthesis of Fe-doped $MoS_2$ monolayer**: $MoO_3$ (15 mg) and $FeS_2$ (99.99%, Aldrich) were mixed at a molar ratio of Fe to Mo of 0.12, 0.24, 0.48 and 0.4 and placed in a quartz boat inside a quartz tube at the center of a heating zone. A clean $SiO_2$/Si substrate (285 nm $SiO_2$) was treated with PTCA solution (5% wt) and placed in the downstream region. The quartz tube was first evacuated to a pressure of 20 mTorr, followed by a 50 sccm flow of Ar/$H_2$ (95%/5%). Sulfur power (1 g) was placed in a separate quartz boat in the upstream region and heated from room temperature to 290 °C by a heating belt when the $MoO_3$ temperature was ~500 °C. The Fe-doped $MoS_2$ monolayer was synthesized by three-step heating process: 20-600 °C for 30 min, 600-760 °C for 40 min, and 760 °C for 15 min. Finally, all heating was stopped and the samples were cooled to room temperature.

**Optical Spectroscopy Measurement:** The PL and Raman signals were measured by a Witec Alpha 300R confocal Raman microscope system, coupled with a 7 T superconductive magnetic field and a closed cycle optical cryostat (10 K). In the polarization-resolved PL measurement system, a linearly polarized laser beam passes through a $1/4\lambda$ waveplate, and is then focused on the samples by a long working distance objective (Olympus, 50x, NA = 0.45). The circularly polarized PL is collected by the same objective, and passes through the same $1/4\lambda$ waveplate and an analyzer into a spectrometer and a CCD camera (see Supplementary Information Fig. S12). The PL and Raman spectra were collected under a 2.41 eV laser excitation of 3.0 mW. The overlap of $MoS_2$ and Si Raman features with photoluminescence features can conceal the splitting under nearly on-resonance excitation (1.96 eV), especially at low temperature (10 K) (see Supplementary Information Fig. S13); therefore, the excitation wavelength was detuned to 2.41 eV in our experiments.

**STEM Characterization and Image Simulation**: Fe-doped monolayer $MoS_2$ was transferred onto a TEM grid from the $SiO_2$/Si substrates using the PMMA-assisted method. STEM-ADF imaging was performed by an aberration-corrected JEOL ARM-200F with a cold field emission gun and an ASCOR probe corrector at 60 kV. The convergence semiangle of the probe was ≈30 *mrad*. STEM-ADF images were collected using a half-angle range from ≈85 to 280 *mrad*. Dwell times of 19 and 10 μs pixel$^{-1}$ were set for single-scan imaging and sequential imaging, respectively. The QSTEM package was used for image simulations by assuming an aberration-free probe and an ≈1 Å source size to give a probe size of ≈1.2 Å.

**Associated content**

**Supporting Information**

The authors declare no competing financial interest.

The Supporting Information is available free of charge on the ACS Publications website

1. Thickness of Fe-doped monolayer $MoS_2$.

2. Large-area atomic-resolution STEM-ADF image of CVD-grown Fe-doped monolayer $MoS_2$.

3. EELs mapping of CVD-grown Fe-doped monolayer $MoS_2$.

4. Optical system calibration by polarization-resolved Si Raman spectra.

5. Reflective magnetic circular dichroism measurement (RMCD).

6. Control experiments in undoped $MoS_2$ monolayer.

7. Valley Zeeman splitting increase with increasing Fe dopant concentration.

8. Excluding the possibility of carrier doping and defects for $g_{eff}$ factor enhancement in Fe-doped monolayer $MoS_2$.

9. Room-temperature valley Zeeman splitting in Fe-doped monolayer $MoS_2$.

10. Calculation details and results of Fe-doped monolayer $MoS_2$.

11. The density of states (DOS) of Fe-doped $MoS_2$ monolayer calculated from a 5x5x1 $MoS_2$ supercell with one Fe atom substituting one Mo atom.

12. Confocal polarization-resolved PL/Raman microscope system.

13. Valley polarization in Fe-doped monolayer $MoS_2$ upon on-resonance excitation.

**Corresponding Author**

* Bo Peng: bo_peng@uestc.edu.cn

* Kian Ping Loh: chmlohkp@nus.edu.sg

* Yingchun Cheng: iamyccheng@njtech.edu.cn

**ORCID**

Bo Peng: 0000-0001-9411-716X

Kian Ping Loh: 0000-0002-1491-743X

**Author Contributions**

B.P. and L.J.D developed the concept, designed the experiment and prepared the manuscript. Q.L, S.Z.T. performed the polarization-resolved PL measurements. X.X.Z, K.P.L and S.J.P contributed to the measurement of STEM-ADF. M.J.Z, L.B.Z synthesized the monolayer Fe-doped $MoS_2$. S.L., Q.H.X contributed to the study of AFM. Y.C.C., Q.L.W., W.H. contributed to fully relativistic calculations. B.P, W.B.G., L.Z., H.P.L., C.W. Q., and G.X. contributed to the study of mechanism.

**Acknowledgments**

We acknowledge financial support from National Science Foundation of China (51602040, 51872039), Science and Technology Program of Sichuan (M112018JY0025) and Scientific Research Foundation for New Teachers of UESTC (A03013023601007). We thank Prof. Jianhua Zhao and Dr. Hailong Wang from Institute of Semiconductors, Chinese Academy of Sciences for the helpful suggestions on the diluted magnetic semiconductor.

**References**

1. Tombros, N.; Jozsa, C.; Popinciuc, M.; Jonkman, H. T.; van Wees, B. J. Electronic Spin Transport and Spin Precession in Single Graphene Layers at Room

Temperature. *Nature* **2007**, *448*, 571-574.

2. Bonilla, M.; Kolekar, S.; Ma, Y.; Diaz, H. C.; Kalappattil, V.; Das, R.; Eggers, T.; Gutierrez, H. R.; Phan, M.-H.; Batzill, M. Strong Room-Temperature Ferromagnetism in VSe$_2$ Monolayers on van der Waals Substrates. *Nat. Nanotechnol.* **2018**, *13*, 289-293.

3. MacNeill, D.; Heikes, C.; Mak, K. F.; Anderson, Z.; Kormányos, A.; Zólyomi, V.; Park, J.; Ralph, D. C. Breaking of Valley Degeneracy by Magnetic Field in Monolayer MoSe$_2$. *Phys. Rev. Lett.* **2015**, *114*, 037401.

4. Yuan, H.; Bahramy, M. S.; Morimoto, K.; Wu, S.; Nomura, K.; Yang, B.-J.; Shimotani, H.; Suzuki, R.; Toh, M.; Kloc, C.; Xu, X.; Arita, R.; Nagaosa, N.; Iwasa, Y. Zeeman-Type Spin Splitting Controlled by an Electric Field. *Nat. Phys.* **2013**, *9*, 563-569.

5. Aivazian, G.; Gong, Z.; Jones, A. M.; Chu, R.-L.; Yan, J.; Mandrus, D. G.; Zhang, C.; Cobden, D.; Yao, W.; Xu, X. Magnetic Control of Valley Pseudospin in Monolayer WSe$_2$. *Nat. Phys.* **2015**, *11*, 148-152.

6. Xiao, D.; Liu, G.; Feng, W.; Xu, X.; Yao, W. Coupled Spin and Valley Physics in Monolayers of MoS$_2$ and Other Group-VI Dichalcogenides. *Phys. Rev. Lett.* **2012**, *108*, 196802.

7. Zeng, H.; Dai, J.; Yao, W.; Xiao, D.; Cui, X. Valley Polarization in MoS$_2$ Monolayers by Optical Pumping. *Nat. Nanotechnol.* **2012**, *7*, 490-493.

8. Peng, B.; Li, Q.; Liang, X.; Song, P.; Li, J.; He, K.; Fu, D.; Li, Y.; Shen, C.; Wang, H.; Wang, C.; Liu, T.; Zhang, L.; Lu, H.; Wang, X.; Zhao, J.; Xie, J.; Wu, M.; Bi, L.;

Deng, L., *et al.* Valley Polarization of Trions and Magnetoresistance in Heterostructures of MoS$_2$ and Yttrium Iron Garnet. *ACS Nano* **2017**, *11*, 12257-12265.

9. Xu, X.; Yao, W.; Xiao, D.; Heinz, T. F. Spin and Pseudospins in Layered Transition Metal Dichalcogenides. *Nat. Phys.* **2014**, *10*, 343-350.

10. Srivastava, A.; Sidler, M.; Allain, A. V.; Lembke, D. S.; Kis, A.; Imamoglu, A. Valley Zeeman Effect in Elementary Optical Excitations of Monolayer WSe$_2$. *Nat. Phys.* **2015**, *11*, 141-147.

11. Stier, A. V.; McCreary, K. M.; Jonker, B. T.; Kono, J.; Crooker, S. A. Exciton Diamagnetic Shifts and Valley Zeeman Effects in Monolayer WS$_2$ and MoS$_2$ to 65 Tesla. *Nat. Commun.* **2016**, *7*, 10643.

12. Wu, Y. J.; Shen, C.; Tan, Q. H.; Shi, J.; Liu, X. F.; Wu, Z. H.; Zhang, J.; Tan, P. H.; Zheng, H. Z. Valley Zeeman Splitting of Monolayer MoS$_2$ Probed by Low-Field Magnetic Circular Dichroism Spectroscopy at Room Temperature. *Appl. Phys. Lett.* **2018**, *112*, 153105.

13. Li, Y.; Ludwig, J.; Low, T.; Chernikov, A.; Cui, X.; Arefe, G.; Kim, Y. D.; van der Zande, A. M.; Rigosi, A.; Hill, H. M.; Kim, S. H.; Hone, J.; Li, Z.; Smirnov, D.; Heinz, T. F. Valley Splitting and Polarization by the Zeeman Effect in Monolayer MoSe$_2$. *Phys. Rev. Lett.* **2014**, *113*, 266804.

14. Jiang, C.; Liu, F.; Cuadra, J.; Huang, Z.; Li, K.; Rasmita, A.; Srivastava, A.; Liu, Z.; Gao, W.-B. Zeeman Splitting *via* Spin-Valley-Layer Coupling in Bilayer MoTe$_2$. *Nat. Commun.* **2017**, *8*, 802.

15. Zhao, C.; Norden, T.; Zhang, P.; Zhao, P.; Cheng, Y.; Sun, F.; Parry, J. P.; Taheri, P.; Wang, J.; Yang, Y.; Scrace, T.; Kang, K.; Yang, S.; Miao, G.-X.; Sabirianov, R.; Kioseoglou, G.; Huang, W.; Petrou, A.; Zeng, H. Enhanced Valley Splitting in Monolayer WSe$_2$ due to Magnetic Exchange Field. *Nat. Nanotechnol.* **2017**, *12*, 757-762.

16. Zhong, D.; Seyler, K. L.; Linpeng, X.; Cheng, R.; Sivadas, N.; Huang, B.; Schmidgall, E.; Taniguchi, T.; Watanabe, K.; McGuire, M. A.; Yao, W.; Xiao, D.; Fu, K.-M. C.; Xu, X. van der Waals Engineering of Ferromagnetic Semiconductor Heterostructures for Spin and Valleytronics. *Sci. Adv.* **2017**, *3*, e1603113.

17. Ye, Y.; Xiao, J.; Wang, H.; Ye, Z.; Zhu, H.; Zhao, M.; Wang, Y.; Zhao, J.; Yin, X.; Zhang, X. Electrical Generation and Control of the Valley Carriers in a Monolayer Transition Metal Dichalcogenide. *Nat. Nanotechnol.* **2016**, *11*, 598-602.

18. Wang, Z.; Mak, K. F.; Shan, J. Strongly Interaction-Enhanced Valley Magnetic Response in Monolayer WSe$_2$. *Phys. Rev. Lett.* **2018**, *120*, 066402.

19. Back, P.; Sidler, M.; Cotlet, O.; Srivastava, A.; Takemura, N.; Kroner, M.; Imamoğlu, A. Giant Paramagnetism-Induced Valley Polarization of Electrons in Charge-Tunable Monolayer MoSe$_2$. *Phys. Rev. Lett.* **2017**, *118*, 237404.

20. Sidler, M.; Back, P.; Cotlet, O.; Srivastava, A.; Fink, T.; Kroner, M.; Demler, E.; Imamoglu, A. Fermi Polaron-Polaritons in Charge-Tunable Atomically Thin Semiconductors. *Nat. Phys.* **2016**, *13*, 255.

21. Kang, M.; Jung, S. W.; Shin, W. J.; Sohn, Y.; Ryu, S. H.; Kim, T. K.; Hoesch, M.; Kim, K. S. Holstein Polaron in a Valley-Degenerate Two-Dimensional Semiconductor.

*Nat. Mater.* **2018**, *17*, 676-680.

22. Gustafsson, M. V.; Yankowitz, M.; Forsythe, C.; Rhodes, D.; Watanabe, K.; Taniguchi, T.; Hone, J.; Zhu, X.; Dean, C. R. Ambipolar Landau Levels and Strong Band-Selective Carrier Interactions in Monolayer $WSe_2$. *Nat. Mater.* **2018**, *17*, 411-415.

23. Furdyna, J. K. Diluted Magnetic Semiconductors. *J. Appl. Phys.* **1988**, *64*, R29-R64.

24. Furdyna, J. K. Diluted Magnetic Semiconductors: An Interface of Semiconductor Physics and Magnetism (Invited). *J. Appl. Phys.* **1982**, *53*, 7637-7643.

25. Furdyna, J. K., Kossut, J. Diluted Magnetic Semiconductors. In *Semiconductors and Semimetals*; Willardson, R. K., Beer, A. C., Eds.; Academic Press, inc.: London, 1988; pp 252-257.

26. Furdyna, J. K.; Samarth, N. Magnetic Properties of Diluted Magnetic Semiconductors: A Review (Invited). *J. Appl. Phys.* **1987**, *61*, 3526-3531.

27. Beaulac, R.; Archer, P. I.; Liu, X.; Lee, S.; Salley, G. M.; Dobrowolska, M.; Furdyna, J. K.; Gamelin, D. R. Spin-Polarizable Excitonic Luminescence in Colloidal $Mn^{2+}$-Doped CdSe Quantum Dots. *Nano Lett.* **2008**, *8*, 1197-1201.

28. Gaj, J. A., Kossut, J. Introduction to the Physics of Diluted Magnetic Semiconductors. In *Springer Series in Materials Science*; Hull, R., Jagadish, C., Kawazoe, Y., Kruzic, J., Osgood, R. M., Parisi, J., Pohl, U. W., Seong, T.-Y., Uchida, S.-i., Wang, Z.M., Eds.; Springer, Berlin, Heidelberg, 2010.

29. Yu, J. H.; Liu, X.; Kweon, K. E.; Joo, J.; Park, J.; Ko, K.-T.; Lee, D. W.; Shen, S.;

Tivakornsasithorn, K.; Son, J. S.; Park, J.-H.; Kim, Y.-W.; Hwang, G. S.; Dobrowolska, M.; Furdyna, J. K.; Hyeon, T. Giant Zeeman Splitting in Nucleation-Controlled Doped CdSe:Mn$^{2+}$ Quantum Nanoribbons. *Nat. Mater.* **2009**, *9*, 47.

30. MacDonald, A. H.; Schiffer, P.; Samarth, N. Ferromagnetic Semiconductors: Moving Beyond (Ga,Mn)As. *Nat. Mater.* **2005**, *4*, 195-202.

31. Ohno, H.; Munekata, H.; Penney, T.; von Molnár, S.; Chang, L. L. Magnetotransport Properties of *p*-Type (In,Mn)As Diluted Magnetic III-V Semiconductors. *Phys. Rev. Lett.* **1992**, *68*, 2664-2667.

32. Cheng, Y. C.; Zhu, Z. Y.; Mi, W. B.; Guo, Z. B.; Schwingenschlögl, U. Prediction of Two-Dimensional Diluted Magnetic Semiconductors: Doped Monolayer MoS$_2$ Systems. *Phys. Rev. B* **2013**, *87*, 100401.

33. Ramasubramaniam, A.; Naveh, D. Mn-Doped Monolayer MoS$_2$: An Atomically Thin Dilute Magnetic Semiconductor. *Phys. Rev. B* **2013**, *87*, 195201.

34. Gong, Y.; Liu, Z.; Lupini, A. R.; Shi, G.; Lin, J.; Najmaei, S.; Lin, Z.; Elías, A. L.; Berkdemir, A.; You, G.; Terrones, H.; Terrones, M.; Vajtai, R.; Pantelides, S. T.; Pennycook, S. J.; Lou, J.; Zhou, W.; Ajayan, P. M. Band Gap Engineering and Layer-by-Layer Mapping of Selenium-Doped Molybdenum Disulfide. *Nano Lett.* **2013**, 442-449.

35. Huang, J.-K.; Pu, J.; Hsu, C.-L.; Chiu, M.-H.; Juang, Z.-Y.; Chang, Y.-H.; Chang, W.-H.; Iwasa, Y.; Takenobu, T.; Li, L.-J. Large-Area Synthesis of Highly Crystalline WSe$_2$ Monolayers and Device Applications. *ACS Nano* **2014**, *8*, 923-930.

36. Wang, S.; Wang, X.; Warner, J. H. All Chemical Vapor Deposition Growth of MoS$_2$: h-BN Vertical van der Waals Heterostructures. *ACS Nano* **2015**, *9*, 5246-5254.

37. Najmaei, S.; Liu, Z.; Zhou, W.; Zou, X.; Shi, G.; Lei, S.; Yakobson, B. I.; Idrobo, J.-C.; Ajayan, P. M.; Lou, J. Vapour Phase Growth and Grain Boundary Structure of Molybdenum Disulphide Atomic Layers. *Nat. Mater.* **2013**, *12*, 754-759.

38. Liu, Y.; Nan, H.; Wu, X.; Pan, W.; Wang, W.; Bai, J.; Zhao, W.; Sun, L.; Wang, X.; Ni, Z. Layer-by-Layer Thinning of MoS$_2$ by Plasma. *ACS Nano* **2013**, 4202-4209.

39. Christopher, J. W.; Goldberg, B. B.; Swan, A. K. Long Tailed Trions in Monolayer MoS$_2$: Temperature Dependent Asymmetry and Resulting Red-Shift of Trion Photoluminescence Spectra. *Sci. Rep.* **2017**, *7*, 14062.

40. Peng, B.; Yu, G.; Zhao, Y.; Xu, Q.; Xing, G.; Liu, X.; Fu, D.; Liu, B.; Tan, J. R. S.; Tang, W.; Lu, H.; Xie, J.; Deng, L.; Sum, T. C.; Loh, K. P. Achieving Ultrafast Hole Transfer at the Monolayer MoS$_2$ and CH$_3$NH$_3$PbI$_3$ Perovskite Interface by Defect Engineering. *ACS Nano* **2016**, *10*, 6383-6391.

41. Zhao, X.; Dan, J.; Chen, J.; Ding, Z.; Zhou, W.; Loh, K. P.; Pennycook, S. J. Atom-by-Atom Fabrication of Monolayer Molybdenum Membranes. *Adv. Mater.* **2018**, *30*, 1707281.

42. Guldner, Y.; Rigaux, C.; Menant, M.; Mullin, D. P.; Furdyna, J. K. Magnetooptical Evidence of Exchange Interactions in Zero-Gap Hg$_{1-X}$Fe$_x$Te Mixed Crystals. *Solid State Commun.* **1980**, *33*, 133-136.

43. Mak, K. F.; He, K.; Lee, C.; Lee, G. H.; Hone, J.; Heinz, T. F.; Shan, J. Tightly Bound Trions in Monolayer MoS$_2$. *Nat. Mater.* **2013**, *12*, 207-211.

44. Xiao, D.; Liu, G.; Feng, W.; Xu, X.; Yao, W. Coupled Spin and Valley Physics in Monolayers of MoS$_2$ and Other Group-VI Dichalcogenides. *Phys. Rev. Lett.* **2012**, *108*, 196802.

45. Zhang, Q.; Yang, S. A.; Mi, W.; Cheng, Y.; Schwingenschlögl, U. Large Spin-Valley Polarization in Monolayer MoTe$_2$ on Top of Euo(111). *Adv. Mater.* **2016**, *28*, 959-966.

46. Cheng, Y. C.; Zhang, Q. Y.; Schwingenschlögl, U. Valley Polarization in Magnetically Doped Single-Layer Transition-Metal Dichalcogenides. *Phys. Rev. B* **2014**, *89*, 155429.

47. Serre, H., Bastard, G., Rigaux, C., Mycielski, J., Furdyna, J. K. Physics of Narrow Gap Semiconductors. Infrared Magnetoabsorption in Zero Gap Hg$_{1-X}$Fe$_x$Te and Hg$_{1-X}$Fe$_x$Se Mixed Crystals. In *Physics of Narrow Gap Semiconductors*; Gornik, E., HeinrichL, H., Palmetshofer, L., Eds.; Springer Berlin Heidelberg: 1982; pp 321-325.

48. Bramwell, S. T. Temperature Dependence of the Isotropic Exchange Constant. *J. Phys.: Condens. Matter* **1990**, *2*, 7527.

49. Sevik, C. Assessment on Lattice Thermal Properties of Two-Dimensional Honeycomb Structures: Graphene, h-BN, h-MoS$_2$ and h-MoSe$_2$. *Phys. Rev. B* **2014**, *89*, 035422.

50. Murray, R.; Evans, B. The Thermal Expansion of 2H-MoS$_2$ and 2H-WSe$_2$ between 10 and 320 K. *J. Appl. Crystallogr.* **1979**, *12*, 312-315.

51. Spalek, J.; Lewicki, A.; Tarnawski, Z.; Furdyna, J. K.; Galazka, R. R.; Obuszko,

Z. Magnetic Susceptibility of Semimagnetic Semiconductors: The High-Temperature Regime and the Role of Superexchange. *Phys. Rev. B* **1986**, *33*, 3407-3418.

52. Mermin, N. D.; Wagner, H. Absence of Ferromagnetism or Antiferromagnetism in One- or Two-Dimensional Isotropic Heisenberg Models. *Phys. Rev. Lett.* **1966**, *17*, 1133-1136.

53. Kang, J.; Tongay, S.; Zhou, J.; Li, J.; Wu, J. Band Offsets and Heterostructures of Two-Dimensional Semiconductors. *Appl. Phys. Lett.* **2013**, *102*, 012111.

54. Song, T.; Cai, X.; Tu, M. W.-Y.; Zhang, X.; Huang, B.; Wilson, N. P.; Seyler, K. L.; Zhu, L.; Taniguchi, T.; Watanabe, K.; McGuire, M. A.; Cobden, D. H.; Xiao, D.; Yao, W.; Xu, X. Giant Tunneling Magnetoresistance in Spin-Filter van der Waals Heterostructures. *Science* **2018**, *360*, 1214-1218.

55. Deng, Y.; Yu, Y.; Song, Y.; Zhang, J.; Wang, N. Z.; Sun, Z.; Yi, Y.; Wu, Y. Z.; Wu, S.; Zhu, J.; Wang, J.; Chen, X. H.; Zhang, Y. Gate-Tunable Room-Temperature Ferromagnetism in Two-Dimensional $Fe_3GeTe_2$. *Nature* **2018**, *563*, 94-99.

56. Klein, D. R.; MacNeill, D.; Lado, J. L.; Soriano, D.; Navarro-Moratalla, E.; Watanabe, K.; Taniguchi, T.; Manni, S.; Canfield, P.; Fernández-Rossier, J.; Jarillo-Herrero, P. Probing Magnetism in 2D van der Waals Crystalline Insulators *via* Electron Tunneling. *Science* **2018**, *360*, 1218-1222.

57. Huang, B.; Clark, G.; Navarro-Moratalla, E.; Klein, D. R.; Cheng, R.; Seyler, K. L.; Zhong, D.; Schmidgall, E.; McGuire, M. A.; Cobden, D. H.; Yao, W.; Xiao, D.; Jarillo-Herrero, P.; Xu, X. Layer-Dependent Ferromagnetism in a van der Waals Crystal Down to the Monolayer Limit. *Nature* **2017**, *546*, 270-273.

58. Gong, C.; Li, L.; Li, Z.; Ji, H.; Stern, A.; Xia, Y.; Cao, T.; Bao, W.; Wang, C.; Wang, Y.; Qiu, Z. Q.; Cava, R. J.; Louie, S. G.; Xia, J.; Zhang, X. Discovery of Intrinsic Ferromagnetism in Two-Dimensional van der Waals Crystals. *Nature* **2017**, *546*, 265–269.

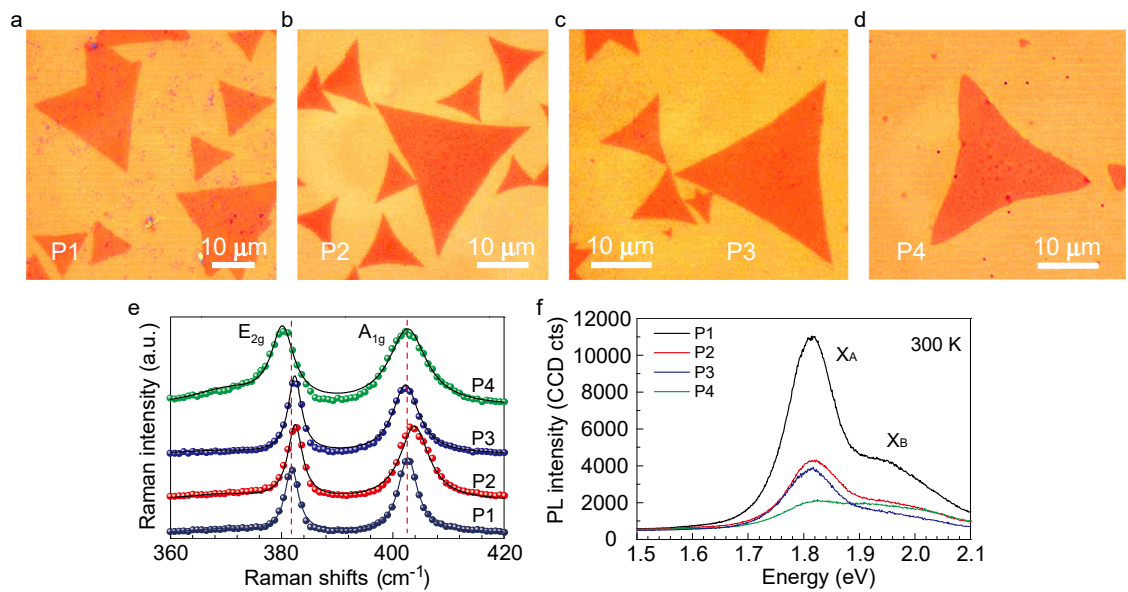

**Figure 1.** Fe-doped MoS$_2$ monolayer. (a-d) Optical microscope images of CVD-grown Fe-doped monolayer MoS$_2$ with the source molar ratio of Fe to Mo of 0.12 (a), 0.24 (b), 0.48 (c) and 0.4 (d). (e-f) Corresponding Raman and photoluminescence spectra of Fe-doped monolayer MoS$_2$ at room temperature. The PL peaks of A (X$_A$) and B (X$_B$) excitons are observed. No trion is induced by the Fe dopants (Fig. S8a-b).

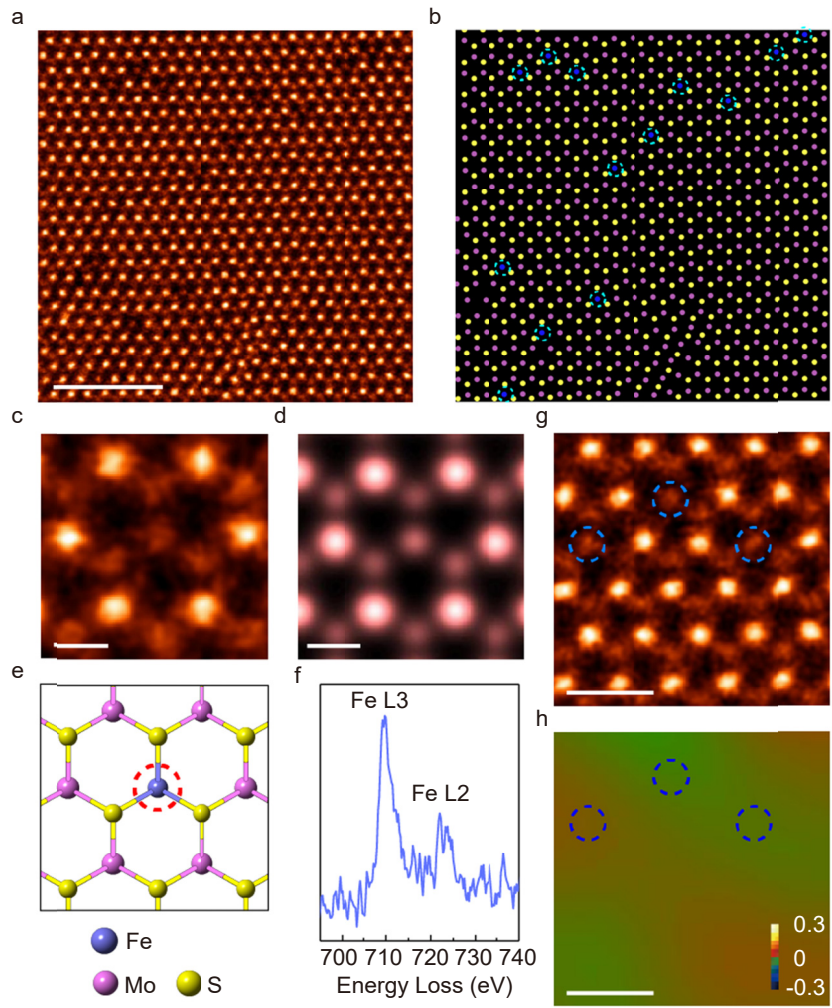

**Figure 2.** Atomic-resolution STEM-ADF images of Fe-doped $MoS_2$ monolayer. (a) Typical STEM-ADF image of Fe-doped 1H-$MoS_2$ monolayer. (b) Corresponding atomic mode of (a) showing the distribution of Fe dopant atoms. Fe, Mo, and S atoms are depicted as blue, pink, and yellow spheres. (c) Enlarged STEM-ADF image revealing a single Fe dopant atom, (d) Corresponding simulated image, and (e) atomic mode. (f) EELs taken from the dopant sites showing a clear Fe L-edge. (g, h) STEM-ADF image (g) containing three Fe dopant atoms (blue dashed circles) and corresponding strain analysis (h), indicating no strain is introduced by Fe dopants. Scale bars: 2 nm in (a), 0.2 nm in (c-d), and 0.5 nm in (g-h).

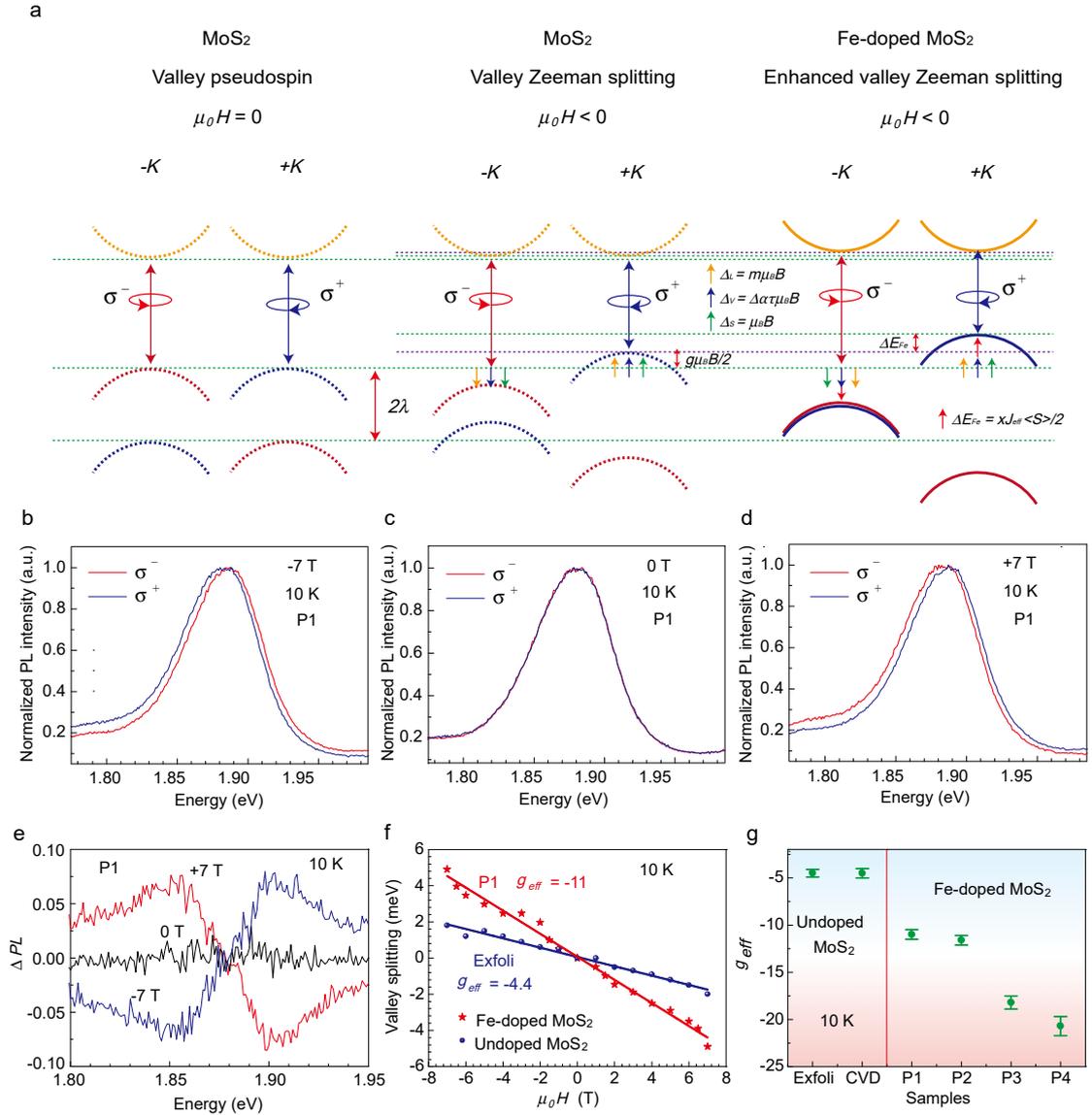

**Figure 3.** Valley Zeeman splitting. (a) Cartoons illustrating the band structures of the undoped and Fe-doped monolayer $MoS_2$ near the $\pm K$ points in the absence and presence of a negative magnetic field ($\mu_0 H$, ↓). The valance band of pristine $MoS_2$ monolayer is splitted by $2\lambda$ due to SOC (Fig. 3a, left). The external magnetic field lifts the valley degeneracy, leading to valley Zeeman splitting (Fig. 3a, middle). Fe dopant enhances the interaction of the magnetic moment and magnetic field, which gives rise to much more pronounced valley splitting (Fig. 3a, right). (b-d) Normalized raw polarization-resolved valley-exciton PL spectra of Fe-doped monolayer $MoS_2$ (P1,

Fig. 1a) under magnetic field of 0 and ±7 T at 10 K. The PL spectra from the +$K$ and –$K$ valleys are completely overlapped at 0 T, but they split at +7 T and –7 T, and the PL energy shifts are opposite. To clearly demonstrate the splitting, only PL peaks of A exciton are shown. (e) Difference between the normalized raw σ⁻ and σ⁺ PL components at 0 and ±7 T, which display no difference at 0 T but obvious opposite trends at ±7 T, manifesting the occurrence of valley Zeeman splitting. (f) Valley splitting as a function of magnetic field in undoped MoS$_2$ (exfoliated) and Fe-doped MoS$_2$ (P1) monolayers at 10 K. The localized Fe spins ($\langle S \rangle$) and intrinsic Zeeman splitting ($g\mu_B B$) are linearly dependent with magnetic field (Ref. 25, 42 and 47), thus valley Zeeman splitting of Fe-doped monolayer MoS$_2$ follow a linear variation as a function of magnetic field. (g) $g_{eff}$ factor in various of undoped and Fe-doped MoS$_2$ monolayers. The $g_{eff}$ factors are clearly larger in the Fe-doped MoS$_2$ monolayer than that in the undoped MoS$_2$ monolayer; and increase with increasing Fe dopant concentration. All the $g_{eff}$ factors are calculated based on the magnetic-field dependence of valley splitting obtained through "center of mass peak analysis" (Ref. 5).

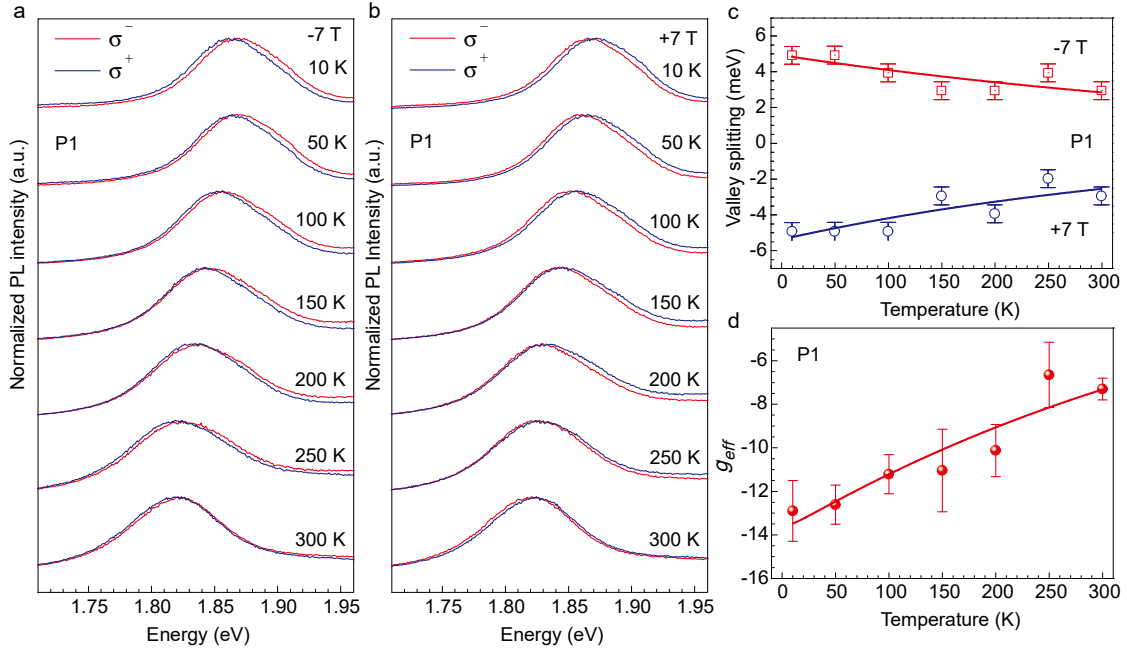

**Figure 4.** Temperature dependence of valley Zeeman splitting. (a, b) Normalized raw polarization-resolved PL spectra of the Fe-doped $MoS_2$ monolayer (P1, Fig. 1a) under a -7 T (a) and +7 T (b) magnetic field as a function of temperature. (c, d) Valley Zeeman splitting (c) and $g_{eff}$ factor (d) as a function of temperature. The solid lines are fits to the data obtained using the temperature dependence of the exchange constant of isotropic Heisenberg exchange interaction mode, as described in the text.

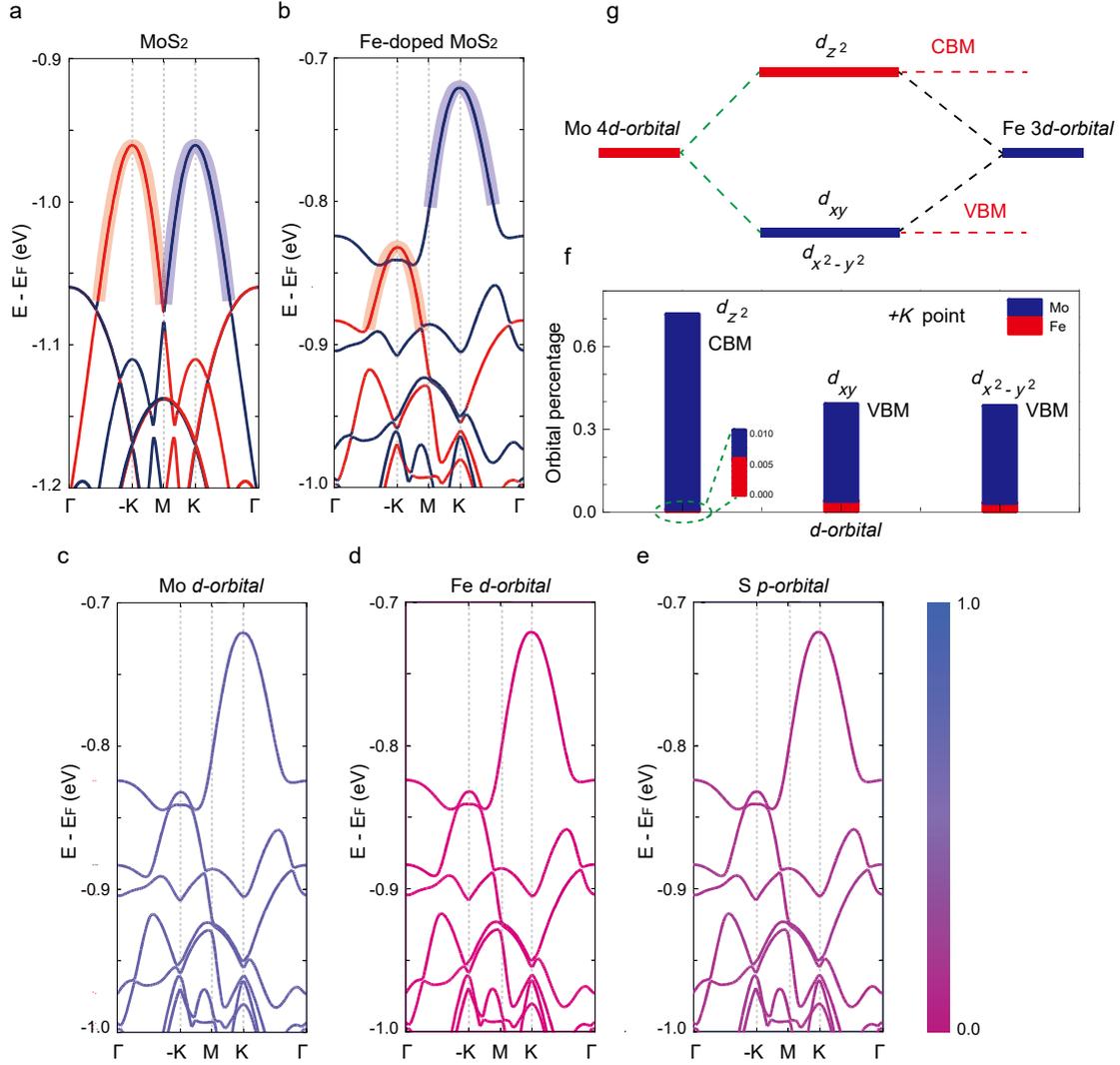

**Figure 5.** Origin of valley Zeeman splitting in Fe-doped monolayer $MoS_2$. (a) Electronic band structures of undoped monolayer $MoS_2$ with SOC, exhibiting an emerging valley pseudospin. (b) Electronic band structures of Fe-doped monolayer $MoS_2$ with SOC and the exchange field, demonstrating valley Zeeman splitting. Blue (red) curves represent spin up (down) states in the $+K$ ($-K$) valley. (c-e) Corresponding electronic band structures of the $d$-orbitals and $p$-orbitals of Mo, Fe and S atoms in Fe-doped monolayer $MoS_2$. The colorbar represents the percentage that each set of orbitals contributes to the overall valence band. (f) Corresponding every isolator $d$-orbital percentage in the hybridized $d$-orbitals of the Fe-doped

monolayer $MoS_2$. The VBM mainly consists of the $d_{x^2-y^2}$ and $d_{xy}$ orbitals of Mo and Fe, while the CBM comprises the $d_{z^2}$ orbitals. The Fe *d*-orbitals unambiguously contribute to the orbital hybridization, although their fractions are tiny as compared to Mo *d*-orbitals. (g) Schematic of the *d*-orbital hybridization of Fe and Mo atoms. All the above calculated results are from a 5 × 5 $MoS_2$ supercell with one Fe atom instead of one Mo atom. The valley Zeeman splitting between the +*K* and –*K* valleys mainly originate from the valance band shifts (Fig. S11b-c), thus only the valance band structures were shown here.

**TOC:**

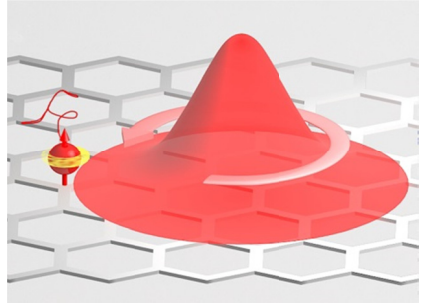